# A Comparative Study on Three Selective Cloud Providers

Rehnuma Tasnim, Afrin Akter Mim,
Salman Hasan Mim, Professor Dr. Md. Ismail Jabiullah

Department of Computer Science & Engineering,
Daffodil International University, Dhaka, Bangladesh

## *Abstract*

*Cloud Computing means a place where we can store our valuable information of data and access the computing and networking services following the pay-as-you-go method without a physical environment. In the present day, cloud computing offers us powerful computing and storage, high availability and security, instant accessibility and adaptation, guaranteed scalability and interoperability, and cost and time effectiveness. Cloud computing has three platforms (IaaS, PaaS, SaaS) with exclusive features which assure to make easy their work for a client, Organization or Trade to build up any kind of IT business. In this paper, we managed a comparison of cloud service features and after the comparison, It's simple to select a certain cloud service from the available features by comparison with three selective cloud providers like Amazon, Microsoft Azure and Digital Ocean. Using the result of this survey to not only find the similarities and differences between various elements of cloud computing but also to propose some topics to look into for further research.*

## *Keywords*

*Cloud Computing, Trending Cloud Providers, cloud Service feature.*

## 1. Introduction

Cloud Computing is being lauded as the next-generation shift that combines the internet and computing, allowing software, material and data to be kept on remote servers that are accessible via the web from anywhere in the world by computers, phones, and televisions, etc. Needless to mention, cloud computing isn't a replacement idea. Indeed, John McCarthy a computer pioneer, predicted, in 1961 that Count would one day be organized as a public-service corporation, and went on to invest in how it could happen. With the passage of your time, we have a brand new have to analyze huge data, thus motivating a greatly increased demand for computing. The field of cloud computing has seen enormous growth in recent years, with a variety of cloud providers emerging. Some providers have concentrated on the computing elements, offering end-users services such as CPU, storage, databases, and networking. Cloud computing is a pricing mechanism that is based on the quantity and duration of services or resources used. Cloud computing offers a wide range of cloud-based services .All clouds have different features, storage capacities, billing systems and different methods to produce the services from other clouds. Differing kinds of service models like Infrastructure-as-a-Service (IaaS), Platform-as-a-service (PaaS) and Software-as-a-service (SaaS) are supported.

All of those approaches had drawbacks, such as high infrastructure costs or inefficient resource consumption. Cloud computing gives people a better option. It consists of big data centers with a collection of tightly coupled resources. The needs of the end-user are met by dynamically regulated resources. Similar to conventional utilities, Consumers can use computing resources from a large pool of resources. The recent problem is that folks don't seem to be aware that which cloud is suitable and consistent with their requirements, they can't ready choose the suitable cloud for or their services among the various clouds managed by different cloud providers. Various cloud computing service companies have emerged over time. Over the last few years, cloud computing has been the subject of significant research, with a range of concerns being debated. Security, privacy, energy management, virtualization, and data management are just a few of the topics being debated. These providers provide various deployment schemes and repair models. Realizing this issue, this research paper conducts a comparative analysis of the current major cloud computing providers. A comparative analysis of various service providers is then provided.

## 2. RELATED WORK

This study conducted completely highlights the attempts from the standpoint of next-generation cellular networking, as well as the open research areas to gain the maximum value from the combination of telecommunication and cloud computing.[2] This paper's comparative results show that the features of each cloud storage system play a significant role in the decision-making process when switching to cloud services.[3] From an economic standpoint, this paper compares several cloud service providers. It computes the costs of services offered by providers in various circumstances.[4] They give a measuring analysis of three main Personal Clouds in this paper: DropBox, Box, and SugarSync. They investigated key components of Personal Cloud storage services in order to define their performance, with a focus on data transmissions. [6] Researchers demonstrated Cloud computing from several perspectives, including definitions, characteristics, and technology. They've depicted a number of representative platforms for Cloud computing.[7]The benefits and drawbacks of cloud computing, cloud storage systems, and infrastructure built using web services like Amazon Web Services are discussed in this article.[8]. They highlighted continuing work on a heterogeneous ocean research platform based on modern decentralized computer and networking architecture.[13]The purpose of this paper is to investigate Microsoft's Windows Azure technology as well as commonly utilized servers.[14]

## 3. AN OVERVIEW OF CLOUD MODEL AND PLATFORM

Simply interpretable that instead of storing a file in a local device or computing in a physical workplace, cloud computing allows you to save any file or access it through the internet and make it secure, reliable and scalable. Basically, the cloud computing model is divided into two categories. They are the development model and the service model, each model also has four categories. Here is a brief of them sequentially.

### 3.1. Cloud Deployment Model

1) Public cloud: It is accessible and usable for everyone. Public clouds operate in the local network and anybody can access them for their regular needs. It is mainly used to connect globally.
2) Private cloud: It's only accessible by a required company or an organization. A private cloud is secure to use and owned by a single organization or a person which is mainly for high-security purposes.
3) Hybrid cloud: Basically it refers to all the functionalities of both public cloud and private cloud into a system named a hybrid cloud. When a company or organization needs all the qualities of both private and public cloud then it's very helpful to use. The private cloud makes the organization secure and the public cloud shares the content and activities of the organization with the customers.
4) Community cloud: When a joint organization needs to exchange its data effectively then it's the perfect solution to exchange their data by using the community cloud. Bank organization is the best example of it.

### 3.1.1. Cloud Service Model

1) Infrastructure as a service (IaaS): IaaS provides the basic computing infrastructure and it's used only for IT administrators by allowing remote servers.
2) Platform as a service (PaaS): PaaS provides a platform where you can develop, test and manage any kind of small or large applications by a software developer.
3) Software as a service (SaaS): Saas provides to hosts and manages the software applications on a pay-as-you-go pricing model. Google drive is the best example of it and it's used by end customers.
4) Unified communication as a service (UCaas): UCaas is based on a technology which is called voice over internet protocol (VOIP). UCAS provides support for enterprise telephone, meetings, unified messaging, instant messaging and presence, mobility, and communications-enabled business processes. It's similar to the software-as-a-service (SaaS) model.

Table 1. A Qualitative and Quantitative Comparison table on Amazon, Azure and Digital Ocean

| Points/Provider Name | Amazon | Azure | Digital Ocean |
| --- | --- | --- | --- |
| Launched | 2007 | 2010 | 2011 |
| Data-Center | 84 | 42 | 12 |
| Services Amount | 100+ | 200+(divided into 18 categories) | 8+ |
| The company take services | Facebook, Netflix, twitch, LinkedIn, BBC, Adobe, Canon, Docker etc | Verizon, MSI Computer, LG electronics, NTT America, Wikimedia Foundation, News Crop, Intel etc | Cerberus-tech, Line-up, Easysize, EMR-bear, Adeva, Hact the box, cloudways, Digital Stage, Obo, JQuery, GitLab, etc |
| Global Market Occupied | 32% | 21% | 8% |
| Service Model | IaaS, PaaS, SaaS | IaaS, PaaS, SaaS | IaaS, SaaS |
| Server OS Type | Any OS | Windows, Linux | Windows, Linux |
| Feature Products | 1. Amazon EC2.(Virtual servers in the cloud) 2. Amazon Simple storage.(Scalable storage in the cloud) 3. Amazon Aurora. (High-performance | 1. Virtual Machines. (Provision Windows and Linux VMs in seconds) 2. Azure Virtual Desktop. (Enable a secure, remote desktop from anywhere) | 1. Droplets. (Scalable virtual machines) 2. Kubernetes. (Managed Kubernetes clusters) 3. App Platform. (Get apps to market faster) 4. Databases. (Worry- |
| | managed relational database)<br>4. Amazon DynamoDB.(Managed NoSQL database)<br>5. Amazon RDS. (Managed MySQL, PostgreSQL, Oracle, SQL Server, and MariaDB)<br>6. AWS Lambda. (Run code without thinking about servers)<br>7. Amazon VPC.(Isolated cloud resources)<br>8. Amazon Lightsail. (launch & manage virtual private servers)<br>9. Amazon SageMaker.(Build,train and deploy machine learning models at scale) | 3. Azure SQL. (Modern SQL family for migration and app modernization)<br>4. Azure Cosmos DB.(Build or modernize scalable & high-performance apps)<br>5. Azure Kubernetes service. (Deploy and scale containers on managed Kubernetes)<br>6. Azure Cognitive services.(add cognitive capabilities to apps with APIs and AI service)<br>7. App Service. (quickly create powerful cloud apps for web and mobile)<br>8. Azure Playfab. (everything you need to build and operate a live game on one platform)<br>9. Azure Functions.(execute event-driven serverless code with an end-to-end development)<br>10. Azure Quantum. (jump in & explore a diverse selection of today's quantum hardware, software and solutions) | free setup & maintenance)<br>5.Spaces.(Simple object storage) |
| Solutions | Archiving, Backup and restore, Blockchain, cloud migration, cloud operations, containers, content delivery, database migrations, data lakes and analytics, DevOps, E-commerce, Edge computing, Front-End web & mobile, High performance computing, Hybrid Cloud Architecture, Internet of Things, Machine learning, Modern application | Application development, Data and Analytics, Security and governance, AI, Hybrid cloud and infrastructure, Cloud migration and modernization, Internet of things, Energy, Financial support, Media and Entertainment, Gaming, HealthCare, Retail, and Manufacturing.<br>*Azure provides 45+ solutions. | Website hosting, Web & Mobile Apps, Video steaming hosting, Gaming development, Cloud VPN, Bid data computing, Startups, SaaS Development, Agency & Web dev shops, Managed cloud hosting providers etc |
| | development, Remote Work, Scientific Computing, Serverless computing, websites, Power and utilities, Semiconductor, sports, etc.<br>*Amazon provides 45+ solutions. | | *Digital Ocean provides 10+ solutions. |
| Supporting Programming Language | Java, python, ruby, PHP, Node Js, Scala Haskell, Perl etc | C#, Java, Node Js, TypeScript ect | Clojure, Go, Java, .NET, iOS, Haskell, Node.js, Perl, python, Ruby, Scala, TypeScript |
| Compute Service | EC2, AWS Lamda, Amazon Lightsail, Elastic BaenStalk | Azure Virtual Machine(VMs), Azure Container Service, Azure App Service, Azure Batch and Azure Service Fabric | Droplets, Kubernetes, App Platform |
| Storage Service | Amazon S3, EBS, S3 Glacier, Flie Storage | Disk Storage, Blob Storage, File Storage, Queue Storage | Block Storage, Space Storage |
| Networking Services | Amazon VPC, Amazon Route 53, Elastic Load Balancing | Azure CDN, Express Route, Virtual Network, Azure DNS | Virtual Private Cloud, Cloud Firewalls, Load Balancer, Floating IPs DNS |
| Type of Load Balancing | Application, Network, Gateway Load Balancer | Azure Load Balancer, Internal Load Balancer(ILB), Traffic Manager | Application, Network, Classic Load Balancer |
| Kubernetes | Elastic Kubernetes Service | Azure kubernetes Service | Digital Ocean Kubernetes (DOKS) |
| Algorithm | The A9 & A10 Algorithm. But A10 Algorithm is very similar to A9. | It has a large library from classification, recommender systems, clustering, anomaly detection, regression and text analytics families for azure machine learning. | Elliptic Curve Digital Signature Algorithm (ECDSA) instead of RSA. |
| Support available | 24/7, Forums, self-help resources, documentation | 24/7, forums, live chats, telephonic communications, documentation | Automated Backups, 24/7 Real-Time Monitoring, Q&A, tutorial, product doc, API doc, etc. |

## 4. CLOUD SERVICE PRODUCT AND COST

Cloud computing provides network, compute and storage services that are simple to use and access over the internet.

### 4.1. Networking Service

Cloud networking service refers to connecting the networking resources as a third-party provider. Every cloud provider has its own networking system and it works in a certain way and is supported by certain things. We focused only on these selective providers' Amazon, Azure and Digital Ocean (As mentioned) networking services. First of all, we know that networking services mainly provide Load balancer, Support IPv4 and IPv6, Virtualization Network, Content delivery network, DNS, Private link and DDos Protection.

In computing, load balancing is a term used to refer to the act of allocating a collection of jobs among a set of resources in order to increase overall processing efficiency. It mainly works to improve the performance of response time. The load balancer monitors the incoming network traffic from the clients and finds the next route request to deliver to the registered destination. It also tracks the suitable destination to ensure the final destination to deliver the resources. There are many types of load balancers which are mentioned below. IP(Internet Protocol) address is also known as a unique address. It is mainly used to identify a device. The difference between IPv4(version 4) and IPv6(version 6) is their size. The size of IPv4 addresses is 32 bits long and IPv6 addresses are 128 bits long.

In the table, all of these three cloud providers provide IPv4 service. For IPv6, Amazon and Azure both supported ipv6 but Digital Ocean does not support ipv6.

Devices can share the same capabilities as a traditional physical network across multiple locations thanks to virtual networking. This feature allows the virtualization method.

The content delivery network feature distributes the networking content in groups where the distribution process is geographically and all of them are connected to servers. It's mainly used for the fast delivery of the content for the clients in different locations which are interconnected with the servers globally. DNS (Domain Name System) is a hierarchical distributed database that lets you store and look up IP addresses and other data by name and Amazon, Azure and Digital Ocean all provide Domain name system services.

In this table, some features are given by the selective cloud providers.

Table 2. A Comparison table on Networking Services

| Features | Amazon | Azure | Digital Ocean |
| --- | --- | --- | --- |
| Load Balancer | Yes | Yes | Yes |
| Types of Load-Balancer | 1.Classic Load Balancer, 2.Network Load Balancer(NLBs), 3.Application Load Balancer(ALBs), 4.Gateway Load | 1.Azure Traffic Manager, 2.Azure Load Balancer, 3.Azure Application Gateway, 4. Azure Front Door | 1.Resize Load Balancer, 2. Scale Load Balancer |

International Journal on Cybernetics & Informatics (IJCI) Vol. 11, No.4, August 2022

|  | Balancer(GWLBs) |  |  |
| --- | --- | --- | --- |
| Support IPv4 & IPv6[2] | Yes | Yes | No (Don't Support IPv6) |
| Virtual Network[3] | Yes | Yes | Yes |
| Content Delivery Network | Yes | Yes | Yes |
| Provide DNS | Yes | Yes | Yes |
| Provide Private Link | Yes | Yes | No |
| DDos Protection | Yes | Yes | No(Cloud Flare) |

## 4.2. Compute Service

Computing Services refers to all information technology and computer systems (including software, application service provider services, hosted computing services, information technology and telecommunication hardware, and other equipment) related to the transmission, storage, maintenance, organization, presentation, generation, processing, or analysis of data and information, whether in an electronic or non-electronic format, whether in an electronic or non-electronic format. We focused on the top three cloud providers in this paper: Amazon, Azure, and Digital Ocean. All of these companies offer a variety of computer services to their clients. As a result, we select one computing service from each of these cloud providers. We use AWS Lamda for Amazon, Azure VM for Azure, and Kubernetes for the Digital Ocean. In this table, we compare these computing services based on a variety of factors such as their billing system, working method, supported languages, and more.

Table 3. A Comparison table on Compute Services

| Features | AWS Lamda | Azure_VM | Kubernetes |
| --- | --- | --- | --- |
| Automatic Scaling | Scaling is done automatically on the size of the workload. It scales the application running the code in response to each trigger. The number of requests that a client's code can process is limitless. Within milliseconds of an incident, AWS Lambda typically starts running your code. Because Lambda scales dynamically, efficiency remains stable as the a number of events grows. | Azure Virtual Machines (VM) is one of several forms of scalable, on-demand computing resources offered by Azure. An Azure VM gives you virtualization flexibility without the need to purchase and maintain the physical hardware that runs it. | To enable autoscaling on an existing node pool, navigate to cluster in the Kubernetes section of the control panel. using Autoscaling in Response to Heavy Resource. A walkthrough that builds an autoscaling cluster and demonstrates the interplay between an HPA and a CA. |

International Journal on Cybernetics & Informatics (IJCI) Vol. 11, No.4, August 2022| | | | |
|---|---|---|---|
| Billing | Pay only for the code running time | Pay only for what you use | Pay-as-you-go payment mechanism |
| Supported languages for computing | C#, Java, python, ruby, PHP, Node Js, Scala Haskell, Perl etc | C#, Java, Node Js, TypeScript , Run time SQL Server, SPA , Oracle ect | Clojure, Go, Java, .NET, iOS, Haskell, Node.js, Perl, python, Ruby, Scala, TypeScript |
| How Does It works | | | |
| Customer | CocaCola, Benchling, Stedi, etc | Sentara, Pearson, Canadian Imperial Bank of Commerce, Accela, Forever21 | Line-Up, Batch, Kea, Adeva, Zuar, CloudWaysetc |
| Service Model | Over 200 AWS Services and SaaS application serverless compute service | PaaS | IaaS |
| Back-up | N/A | Backup Storage is an auto-scaling, reliable set of storage accounts managed by Azure Backup and isolated from customer tenants to provide additional security. Charges for storage are separate from the cost of Azure Backup Protected Instances. | Kubernetes Backup is a data protection solution for vessels assigned as a Kubernetes cluster. Kubernetes backup applications can completely back up all files in containers, incremental backups, or differential backups. It is best to run the Kubernetes backup software frequently and periodically to protect files, configurations, and constantly changing data. |

## 4.3. Storage Service

Traditional network storage and hosted storage are essential to the advancement of cloud storage. The benefit of cloud storage is that you may access your data from anywhere. Cloud storage services can store everything from a single piece of information to an entire company's warehouse. Customers will pay the cloud storage provider based on what they use and how they transmit data to the cloud. Essentially, the cloud storage client replicates the data to any of the cloud storage providers' data servers. Duplicates of data are provided on all or opposite data servers of cloud storage providers so that there is redundancy in availability which ensures that even if something goes wrong the customer's data is protected. Most of the systems store identical data server that uses different power supplies. Because of the rapid development of data and the need to keep it more secure and longer, businesses must integrate how they handle and use the information from the moment the information is created until it is destroyed. We now have the option of storing all of our data on the Internet. Third parties provide and manage this off-site storage over the Internet. Cloud Storage implies that a big pool of storage was available for use with three distinct features: access via the Web Services API over a shaky network connection, quick availability of enormous amounts of storage, and payment for what you use. It supports fast scalability. Cloud storage is an offer of cloud computing. Amazon Web Services, Azure, and Digital Ocean offer many storage tools, but it's not clear which one is best for your needs. Here's the most popular AWS storage are Amazon S3 (Simple Storage Service), Amazon EBS (Elastic Block Store), and Amazon S3 Glacier. Azure storage is Disk Storage, Blob Storage, File Storage, and Queue Storage. And also Digital Ocean storage is Volumes Block Storage, Space Object Storage. We're here to give you an overview of storage, what they're designed for, how they differ, and how to use each service.

Table 4. A Comparison table of Storage services

| Features | AWS | Azure | Digital Ocean |
|---|---|---|---|
| Storage-related Services | Simple Storage Service (S3), Elastic Block Storage (EBS), Elastic File System (EFS), Storage Gateway File Storage | Blob Storage, Queue Storage, File Storage, Disk Storage, Data Lake Storage | Block Storage, Space Object Storage |
| Database-related services | Aurora, DBS, DynamoDB, ElastiCache, Redshift, Neptune, Database Migration Service, | SQL, MySQL, PostgreSQL, Data Warehouse, Server Stretch Database, Cosmos DB, Table Storage, Redis Cache, Data Factory, | MySQL, PostgreSQL, MongoDB Redis, Memgraph, EdgeDB, PhpMyAdmin |
| Backup services | S3 Glacier | Archival Storage Recovery Backups, Site Recovery | Droplet Snapshot (**snapshot**-based **backup** system) |

Table 4.1. A Comparison table of Storage services

| Features | Amazon | Azure | Digital Ocean- |
|---|---|---|---|
| Disk Storage | Yes | Yes | No |
| Data Backup | Yes | Yes | Yes |

| Features | Amazon | Azure | Digital Ocean- |
|---|---|---|---|
| Store and access unstructured data | Yes | Yes | Yes |
| File storage | Yes | Yes | No |
| Data transfer and edge compute | Yes | Yes | Yes |

## 4.4. Pricing Structure

In this table, we examine pricing types and the price differences for different types of virtual CPUs for these suppliers.

Table 5. A Comparison table on pricing type

|  | Amazon | Azure | Digital Ocean |
|---|---|---|---|
| Pricing Type | Pay-as-you-go, On-demand | Pay as you go, on-demand per second billing | Pay as you go pricing. |

Table 6. A Comparison table on basic virtual machine cost

| Virtual CPU | Amazon | Azure | Digital Ocean |
|---|---|---|---|
| 1 vCPU | 8.50$ | 7.59$ | 5$ |
| 2 vCPU | 15.23$ | 15.11$ | 15$ |
| 4 vCPU | 60.91$ | 60.74$ | 40$ |
| 8 vCPU | 121.50$ | 121.18$ | 80$ |

## 5. ANALYSIS OF THE COMPARISON ON AMAZON, AZURE AND DIGITAL OCEAN

The analysis of the above comparisons shows the following result: Amazon and Azure deliver Infrastructure as a Service (IaaS), Platform as a Service (PaaS), and Software as a Service (SaaS), whereas DigitalOcean provides Infrastructure as a ser Service (IaaS). AWS and Azure are better for large scalable applications, whereas Digital Ocean is better for developers and tiny apps. Digital Ocean, mainly suitable for small developers and modest enterprises who need to rapidly set up a small high-performance instance, is its target market. Even still, when it comes to VM performance on the two platforms, Amazon is being beaten by the scrappy rival. Each provider is the best for a variety of different tasks. But overall DigitalOcean is best in terms of VPS Performance based on network speed, CPU usage, web server capacity, CPU-intensive operation rate, and Sequential read/write rates in our selective three cloud providers. [18]

So simply says that "Digital Ocean is not really an Amazon and Azure competitor". But in their continuous triangle pricing battle, Amazon, Microsoft Azure, and Digital Ocean are as formidable as ever. If you expect a lot of data outflow, you should search for companies that offer a substantial provision rather than those that charge by the GB. This is particularly true when data transmission is not interzonal. Unfortunately, all three cloud companies fail to convey relevant subscription information in a logical manner, resorting to price structure obfuscation that borders on nonsense (though Microsoft is a bit better than Amazon and Digital Ocean in this regard). And, while their cloud products may differ in various ways, Amazon's information page appears positively byzantine when contrasted with Digital Ocean's no-nonsense approach.

## 6. FUTURE WORK

This paper enhances in our understanding of cloud services. Despite the fact that we did not discuss the security provided by these cloud providers. In the future, we plan to expand our analysis to include more elements, such as how service providers replicate data and identify some of the top security dangers associated to cloud data, with data loss being the most serious security concern. Cloud computing will be one of the most in-demand occupations in software development, with implications for cloud infrastructure security. Industry lacks the necessary skills to ensure integrity, which is a serious worry. The shortage of trained cyber security specialists is a moderate to significant worry for 93 percent of business.[17]

## 7. CONCLUSIONS

This paper provides an overview of the cloud functionalities provided by major service providers. We compare the most popular cloud providers, including Amazon, Azure, and Digital Ocean. This document explains the many sorts of disparities between these providers in terms of various attributes. This study focuses on the primary services provided by various cloud providers, such as storage, computation, and network services. Data storage, servers, databases, networking, and software, as well as other tools and applications. are some examples of the resources. In recent years, any commercial organization has shifted its operations to the cloud, which has shown to be profitable and attracted the interest of many others. The information gathered in this research paper will help cloud customers choose the significant cloud provider according to their needs as well as the services provided by the selected cloud provider.

### ACKNOWLEDGEMENTS

We, the authors of this paper, would like to express our gratitude to our mentor, for his assistance in completing this research. And thanks to my university Daffodil International University, Bangladesh for support.

## AUTHORS

**Rehnuma Tasnim**. Currently, is pursuing her B.S.C in Computer Science in Computer Science and Engineering from Daffodil International University, Dhaka, Bangladesh. She and her team presented their first paper at the International Conference on IoT, Cloud and Big Data. Her areas of research are Computer networks, cloud infrastructure, IoT, Machine learning, and Big Data.

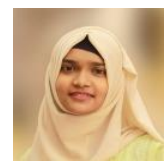

**Afrin Akter Mim**. Currently, is pursuing her B.S.C in Computer Science in Computer Science and Engineering from Daffodil International University. She and her team presented this paper at the International Conference on IoT, Cloud and Big Data (IOTCB 2022). Her areas of research are Computer networks, cloud infrastructure, Machine learning, and Deep Learning.

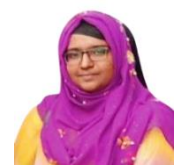

**Salman Hasan Mim** is pursuing B. Sc in Daffodil International University. His strength is attitude and he likes to take on challenges. He and his team presented this paper at the International Conference on IoT, Cloud and Big Data (IOTCB 2022). His areas of research are Computer Networks, Network Security, IoT, Operating Systems and Distributed Computing.

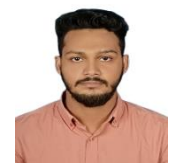

**Professor Dr. Md. Ismail Jabiullah** is currently a Professor at the Department of computer science engineering, Daffodil International University, Dhaka, Bangladesh. He received his B.sc(Hons) and M.sc in Mathematics with first-class from Dhaka University and also an honors scholarship from Dhaka University. He obtained a Ph.D. in Computer science and engineering. His research area is Network Security, Web Security, Software Security, Internet Security, Image Processing, Computer Vision, Wireless Network, Cellular Network, Satellite Network, Artificial Intelligence and Neural Networks, Software Testing, Machine Learning, and Deep Learning. He has published 27 books 78 International journal papers and 105 conference papers.

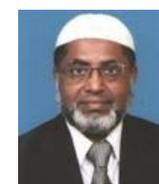